# SINGLE MR IMAGE SUPER-RESOLUTION USING GENERATIVE ADVERSARIAL NETWORK

Shawkh Ibne Rashid
*Faculty of Science, Ontario Tech University*
*Oshawa, Ontario, Canada*
shawkhibne.rashid@ontariotechu.net

Elham Shakibapour
*Faculty of Science, Ontario Tech University*
*Oshawa, Ontario, Canada*
eshakibapour@gmail.com

Mehran Ebrahimi
*Faculty of Science, Ontario Tech University*
*Oshawa, Ontario, Canada*
mehran.ebrahimi@ontariotechu.ca

**ABSTRACT**

Spatial resolution of medical images can be improved using super-resolution methods. Real Enhanced Super Resolution Generative Adversarial Network (Real-ESRGAN) is one of the recent effective approaches utilized to produce higher resolution images, given input images of lower resolution. In this paper, we apply this method to enhance the spatial resolution of 2D MR images. In our proposed approach, we slightly modify the structure of the Real-ESRGAN to train 2D Magnetic Resonance images (MRI) taken from the Brain Tumor Segmentation Challenge (BraTS) 2018 dataset. The obtained results are validated qualitatively and quantitatively by computing SSIM (Structural Similarity Index Measure), NRMSE (Normalized Root Mean Square Error), MAE (Mean Absolute Error) and VIF (Visual Information Fidelity) values.

## 1. INTRODUCTION

Higher quality Magnetic Resonance Images (MRI) are valuable for early detection and accurate diagnosis of various medical conditions. High spatial resolution of images is essential to provide detailed anatomical information and help radiologists with accurate quantitative analysis. Acquiring higher resolution MRI requires higher image acquisition times that can be costly and may not always be possible due to physical limitations. Super-resolution (SR) techniques are alternative ways to improve the spatial resolution of images by producing a High-resolution (HR) image given a Low-resolution (LR) one.

SR approaches to produce HR MRI are mostly categorized into reconstruction-based and learning-based methods (Van et al, 2012). Reconstruction-based techniques use interpolation filtering methods such as bilinear, bicubic or lanczos (Duchon, 1979). They are among the first methods to tackle single image super resolution (SISR) problems. However, interpolated images blur or degrade important edge and texture information of images.

SISR methods have been widely advanced by the breakthroughs in deep learning. Methods based on Generative Adversarial Networks (GANs) (Goodfellow et al, 2014) are promising approaches for image generation and have been also used for SR (Ledig et al, 2017). GANs-based models show the increasing performance for SISR. Different architectures and loss functions aimed at improving the quality of the generated images using GANs have been proposed (Metz et al, 2015, Arjovsky et al, 2017, Mao et al, 2017).

Recent advances in the Super Resolution Generative Adversarial Network (SRGAN) are aimed to recover fine texture details and edge information even at large upscaling factors (Ledig et al, 2017). This motivated us to apply a recent extension of the method called Real Enhanced Super-Resolution Generative Adversarial Network (Real-ESRGAN) that achieves high perceptual quality for 2D real-world images (Wang et al, 2021). Our specific focus in this work is to apply Real-ESRGAN (Wang et al, 2021) to resolution enhancement of 2D slices of 3D MR images. To the best of our knowledge, no studies have been conducted on the use of Real-ESRGAN (Wang et al, 2021) to validate SISR on MRI scans.

## 2. RELATED WORKS

The state-of-the-art methods with deep learning techniques have shown an increasing performance on producing SISR on 2D real-world images (Kim et al, 2016, Lai et al, 2017, Lim et al, 2017, Tai et al, 2017, Haris et al, 2018, Wang et al, 2018, Zhang et al, 2018, Dai et al, 2019, Wang X. et al, 2019). Furthermore, many of the current deep learning techniques for medical image enhancement typically rely on GANs. GANs-based models generate more realistic images (Tan et al, 2020). To produce SISR, a GAN pipeline usually consists of a single generator network that takes in the degraded/down-sampled LR image as input and directly outputs the reconstructed SR image. A photo-metric loss is calculated between the SR image and the ground truth and drives the network to recover realistic image details (Wang J. et al, 2019). Most of the proposed approaches apply SRGAN developed by Ledig et al (2017).

mDCSRN (Chen et al, 2018), Lesion-focussed GAN (Zhu et al, 2019), and ESRGAN (Bing et al, 2019) are GAN-based solutions tackling SR for medical images. In (Tan et al, 2020), a meta-upscale module proposed by Hu et al. (2019) is combined with SRGAN to create a network called Meta-SRGAN. GAN-based models including ESRGAN and CycleGAN are used in (Do et al, 2021) to generate HR MRI with rich textures. Their experimental results have been conducted on both 3T and 7T MRI in recovering different scales of resolution. The authors in (Sanchez et al, 2018) have applied the SRGAN-based model (Ledig et al, 2017) adopted to 3D convolutions to generate HR MRI scans. They have explored different methods for the upsampling phase to alleviate artifacts produced by sub-pixel convolution layers. Chen et al. (2018) have applied the Densely Connected SR Network (DCSRN) (Huang et al, 2016) for 3D brain MR image enhancement. Though, direct conversion into 3D may result in many parameters and thus faces challenges in memory allocation.

SRGAN (Ledig et al, 2017) applies a perceptual loss using high-level feature maps of the VGG network (Simonyan, and Zisserman, 2015, Sprechmann, and LeCun, 2016, Johnson et al, 2016) combined with a discriminator that encourages solutions perceptually difficult to distinguish from the HR reference images. However, the discriminator requires a more powerful capability to discriminate realness from complex training outputs, while the gradient feedback from the discriminator needs to be more accurate for local detail enhancement (Wang et al, 2021).

In Real-ESRGAN (Wang et al, 2021), the VGG-style discriminator in ESRGAN is developed via U-Net design along with spectral normalization (Ronneberger et al, 2015, Schonfeld et al, 2020) to increase discriminator capability and stabilize the training dynamics. As a result, Real-ESRGAN (Wang et al, 2021) achieves better visual performance making it more practical in real-world applications. This motivates us to apply and validate Real-ESRGAN (Wang et al, 2021) to produce SISR of 2D slices of 3D MR images. Real-ESRGAN (Wang et al, 2021) makes use of the RGB LR images. In our proposed approach, we modify the structure of the Real-ESRGAN to train the 2D MR images. The obtained results are assessed and compared quantitatively and qualitatively with the standard bilinear and bicubic interpolation methods.

## 3. METHODOLOGY

In this section, we explain the GAN model we have used to enhance the resolution of brain MRI images, along with the description of the dataset and the modifications we have made to the GAN model for our purpose of working with grayscale brain MR images, cost functions used and the parameter settings.

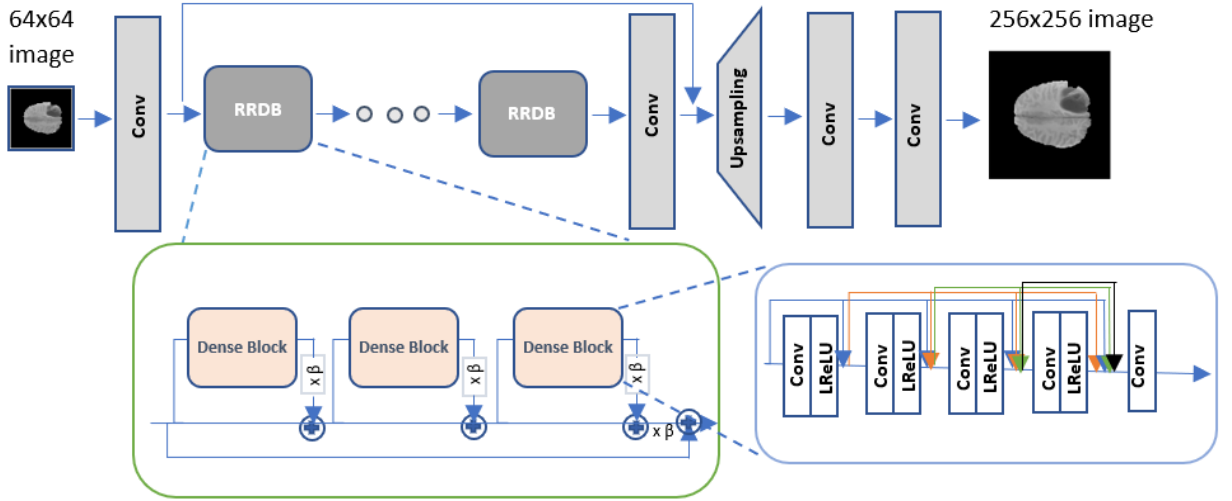

Figure 1. Generator Architecture of Real-ESRGAN.

## 3.1 Dataset Preparation

We have used the BraTS 2018 dataset for the resolution enhancement experiment. The dataset consists of MRI scans of glioblastoma (GBM/HGG) and lower grade glioma (LGG) as native (T1), post-contrast T1 weighted (T1Gd), T2-weighted (T2) and T2 Fluid Attenuated Inversion Recovery (FLAIR) volumes. For the image resolution enhancement purpose, we have used both HGG and LGG T1 brain MR images. There is a total of 285 3D MR T1 weighted images, each consisting of 155 2D slices. The slices are grayscale images and are available in NIfTI format.

We have used 80% of the 3D images for training purpose and the rest of the 20% images for testing the GAN models. For dataset preparation, we have removed the blank MRI scans and have normalized the pixel values to lie in the range of 0 to 1. As this dataset is primarily created for segmentation, the dataset also contains the segmentation masks of the brain MR images. As our purpose is resolution enhancement, we have produced the low-resolution images by down-sampling the 2D MRI scans by a factor of four using bilinear interpolation method. There are some 2D slices that includes no brain tissue.

We have removed such blank MR images. The original size of the 2D images is 240x240. We use zero padding to change the size to 256x256 for training our models. For training, we get 285x0.8x155 = 35340 2D images and after excluding blank images, we have a total of 31322 images. There are a total 7823 images in the test set after the pre-processing steps.

Since we are using a resolution enhancement scale of 4 for low resolution image generation, our low-resolution images are of size 64x64.

## 3.2 Real-ESRGAN Architecture

Real-ESRGAN is a GAN with a Residual in Residual Dense Block (RRDB) based generator and UNET based discriminator. The generator consists of multiple RRDB blocks. This RRDB block is modified from residual block in SRGAN. Instead of using convolutional layers in Residual Blocks, they have used dense blocks (Huang et al, 2016). Each dense block consists of five convolutional layers accompanied with Leaky-RELU layers except for the last one. They have used residual scaling (Szegedy et al, 2016) in residual blocks where output from residual blocks is scaled down by multiplying them with a scale factor in range 0 to 1 (denoted as β in Figure 1). This helps when network becomes deeper and produces very large or small gradients. The RRDB blocks are followed by an upsampling block and two convolutional layers for the reconstruction. For the discriminator, they have used a U-Net based model with skip connections. To stabilize the training dynamics, they used spectral normalization regulation. They have also employed relativistic discriminator based on relativistic generator (Jolicoeur-Martineau, 2018).

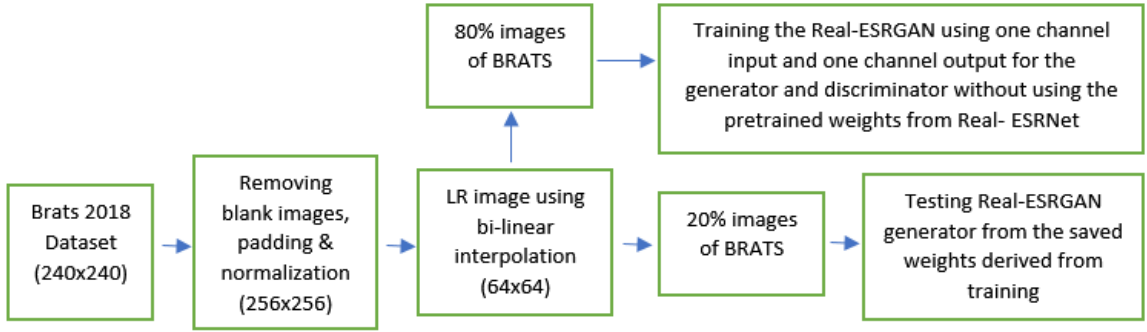

Figure 2. Summary of the training procedure.

### 3.3 Loss Functions

We have used a combination of pixel loss (L1 loss), perceptual loss (Johnson et al, 2016), and GAN loss (Ledig et al, 2017) for training the generator and GAN loss for the discriminator. We have used a weight value of 1 for different losses meaning that we took the total loss values and combined them for training the generator. For discriminator, GAN loss was measured for both fake (generated from generator) and real images (ground truth) so that the discriminator can learn to distinguish between real and fake images. VGG19 weights are used for calculating perceptual loss.

### 3.4 Network Training

According to the Real-ESRGAN paper the training process is divided into two stages. The first part is training Real-ESRNet (Wang et al, 2021) with the L1 loss. When they trained Real-ESRGAN, they used the weights from the Real-ESRNet as the initialization point. In our experiment, we started our training by training the Real-ESRGAN without using the weights from the Real-ESRNet. The pre-trained weights from Real-ESRNet were obtained from training the model with real-world images. As we are working with brain MRI grayscale images, we started our training from scratch using only Real-ESRGAN.

Real-ESRGAN has three channel input and three channel output for the generator and three channel input for the discriminator. As brain MRI data is grayscale, we changed the input and output channel number to 1 for the generator and input channel number to 1 for the discriminator. We first train the model by copying the same MR image three times to match with the original input shape of the model and noticed variations among the three generated images from the generator (for three channel output). As a result, we changed the input shape and output shape of the model to facilitate one single channel training for the grayscale images.

### 3.5 Parameter Settings

We have used 23 Residual in Residual Dense Blocks (RRDB) in the generator. Each residual block has three dense blocks and five convolutional layers in each dense block. The initial input number of channels are set to 64 for these convolutional layers. Channels for each growth for the convolutional layers in the dense block is 32. The growth channel means the first convolutional layer will output 32 features, the second conv layer will output 32x2 = 64 features, the third one will produce 32x3 = 96 features and so on. In the upsampling block, there are two upsampling layers, each increasing the number of features by a factor of two using nearest interpolation method.

We have optimized the generator and discriminator weights using Adam optimizer with a learning rate of 0.0001 based on the loss function as described above. We have trained the Real-ESRGAN for 300000 iterations on the 80% of the BraTS dataset images. A summary of the training procedure is illustrated in Figure 2.

Table 2. The quantitative comparison

| Method | SSIM | NRMSE | MAE | VIF |
|---|---|---|---|---|
| **Bilinear Interpolation** | 0.92±0.03 | 0.04±0.01 | 0.011±0.004 | 0.55±0.07 |
| **Cubic Interpolation** | 0.93±0.03 | 0.04±0.01 | 0.010±0.004 | 0.64±0.07 |
| **Real-ESRGAN** | **0.94±0.03** | 0.04±0.01 | **0.009±0.004** | **0.71±0.09** |

*Note: SSIM, NRMSE, MAE and VIF values of Bilinear Interpolation, Cubic Interpolation, and Real-ESRGAN generated high resolution images compared with ground truth images. The metric values are the mean values for 7823 images. We have also included the standard deviation for these metric values in the table as well. In terms of the metrices, higher value of SSIM and VIF and lower values of NRMSE and MAE denote better quality images.*

## 4. EXPERIMENTAL RESULTS

We have compared the performance of Real-ESRGAN for the MR image resolution enhancement on BraTS 2018 dataset with bilinear and bicubic interpolation methods based on a variety of evaluation metrics including Structural Similarity Index (SSIM), Normalized Root Mean Square Error (NRMSE) (normalized based on the mean value of the ground truth image), Mean Absolute Error (MAE) and Visual Information Fidelity (VIF) (Sheikh, and Bovik, 2006). We have made qualitative and quantitative comparison among these different approaches.

We have run our experiments on a Linux based operating system with two 2199 MHZ processors, 13 GB RAM. We use a NVIDIA Tesla K80 GPU with 12 GB memory. We have used Python programming language and its libraries including OpenCV and PyTorch.

To compare Real-ESRGAN, and different interpolation methods, we have used 20% of the BraTS 2018 dataset. After removing the blank MR images, we obtain 7823 images for testing. We train the Real-ESRGAN for 300000 iterations with a batch size of one. The quantitative comparison can be seen in Table 1.

From Table 1, we can see that Real-ESRGAN produces MR images with higher resolution than other compared methods. Qualitatively speaking, we can see the same thing. Figure 3 shows some of the examples from the test set along with the produced higher resolution images from Bilinear Interpolation, Cubic Interpolation, and Real-ESRGAN.

## 5. CONCLUSION

The problem of generating and validating a single MR Image Super-resolution using Generative Adversarial Network was addressed in this paper. We utilized the Real-ESRGAN model on 2D MR Images available on the BraTS dataset to generate 2D HR MR Images. We demonstrated that the Real-ESRGAN model outperformed the bilinear and bicubic interpolation methods in restoring a resolution by a factor of 4. The generated images were perceptually and qualitatively superior compared to the interpolated ones.

From the obtained results, it can be observed that the interpolated images can be blurry with ghosts and shadows around the boundaries with suppressed sharp edge information. Future work will involve extending the model to generate the SR images with arbitrary zooming factors.

## ACKNOWLDGEMENTS

This work was supported in part by the Natural Sciences and Engineering Research Council of Canada (NSERC).

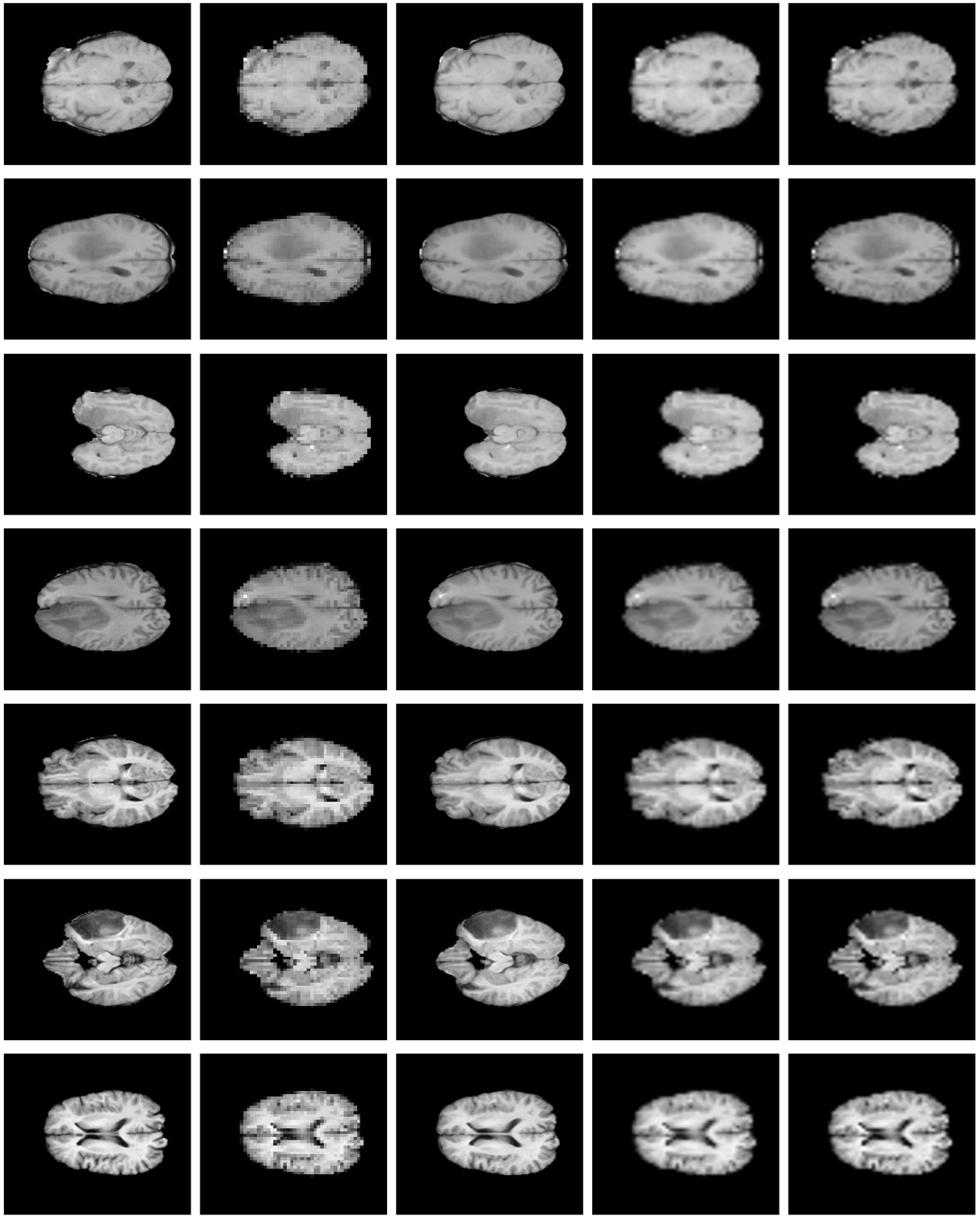

Figure 3. From left to right, Ground truth, Low-resolution image, Generated image based on the model, Bilinear interpolation, and Bicubic interpolation.